\documentstyle[12pt]{article}
\catcode`\@=11

\@addtoreset{equation}{section}
\@addtoreset{equation}{subsection}
\def\theequation{\ifnum\value{subsection}>0\relax
\thesubsection.\arabic{equation}\relax
\else\ifnum\value{section}>0\relax
\thesection.\arabic{equation}\relax
\else\arabic{equation}\fi\fi}

\newcommand{\ba}{\begin{eqnarray}}   
\newcommand{\ea}{\end{eqnarray}}     
\newcommand{\be }{\begin{equation}}   
\newcommand{\ee }{\end{equation}}     

\newcommand{\RMA}{{\cal R}}     

\newcommand{\si}{\sigma}

\begin{document}
\rightline{  }
\vskip 1cm
\centerline{\Large \bf On the Integrability of the One-Dimensional} 
\vskip 0.8cm
\centerline{\Large \bf Open XYZ Spin Chain}

\vskip 0.9cm
\centerline{\bf Guo-xing JU$^{1,2}$, Chi Xiong$^{1}$ }

\vskip0.5cm
\centerline{\sf 
1.Institute of Theoretical Physics,Academia Sinica,Beijing 100080}
                       
\centerline{\sf   
2.Physics Department,   
Henan Normal University,
Xinxiang,Henan Province 453002}                                                       
\vskip0.5cm
\begin{abstract}
The Lax pair for the one-dimensional open XYZ spin chain is constructed, this shows that the system is completely integrable . 
\end{abstract}
\vskip0.5cm

In the formalism of the quantum inverse scattering method
(QISM), the integrability for the systems with periodic boundary conditions is 
guaranteed by the Yang-Baxter equation, of which the central ingredient is {\RMA}-matrix
\cite{fad,kor}. However, it is interesting to find  other boundary conditions 
compatible with integrability for a given model. Based on the work of Cherednik
\cite{che}, Sklyanin\cite{skl} found that there exists boundary condition defined 
by reflection matrices $K^{\pm}$, which satisfy the so-called reflection 
equations, compatible with the Yang-Baxter equations. These $K$-matrices imply 
boundary terms for the Hamiltonians of the spin systems on the finite interval.
Recently,  a lot of integrable models with open boundaries, both in quantum field theory 
and in statistical mechanics, have been studied extensively using this method
\cite{che}-\cite{gon}.\\

On the other hand, for the case of periodic boundary condition, there is  a equivalent 
approach, which is called  the Lax representation \cite{kor}, to the proof of 
the integrability of the system. In this formalism, a model is said to completely
integrable if we can find a Lax pair such that the Lax equation is equivalent 
to the equation of motion of the model. Now, it is natural that one expect a 
variant of the Lax representation to systems with open boundary conditions. In fact, this has been
done, for example, for the one-dimensional (1D) Heisenberg open $XXZ$  chain
\cite{zhou}, the Hubbard model\cite{guan}. In this letter, we will construct the Lax pair for the 1D open $XYZ$ spin-chain.\\

The 1D open $XYZ$ spin-chain has been investigated from the view of the 
reflection equation\cite{vega,ina}. The corresponding Hamiltonian for the system 
with  $N$ sites is given by\cite{ina}
\ba
H &=&-\frac{1}{2}\sum_{n=1}^{N-1}(J_x\si _{n}^{x}\si _{n+1}^{x}+J_y\si _{n}^{y}
\si _{n+1}^{y}+J_z\si _{n}^{z}\si _{n+1}^{z})   \nonumber\\
& &+{\rm  {\rm  sn}} \eta (A_{-}\si_{1}^{z}+
B_{-}\si_{1}^{+}+C_{-}\si_{1}^{-}+A_{+}\si_{N}^{z}
+B_{+}\si_{N}^{+}
+C_{+}\si_{N}^{-}),
\ea
where
\be
J_x=-2(1+k {\rm sn}^2 \eta), J_y=-2(1-k {\rm sn}^2 \eta),J_z=-2{\rm cn} \eta {\rm dn} \eta ,
\ee
$\si_{n}^{x},\si_{n}^{y},\si_{n}^{z}$ are Pauli matrices, and $A_{\pm}, B_{\pm},
C_{\pm}$ are constants. The equations of  motion for the system (1) are given as follows,
\ba
\frac{d}{dt} \si_{n}^x &=& -J_y\si_n^z(\si_{n+1}^y+\si_{n-1}^y)
                      +J_z\si_n^y(\si_{n+1}^z+\si_{n-1}^z),\\
\frac{d}{dt} \si_{n}^y &=& -J_z\si_n^x(\si_{n+1}^z+\si_{n-1}^z)
                      +J_x\si_n^z(\si_{n+1}^x+\si_{n-1}^x),\\
\frac{d}{dt} \si_{n}^z &=& -J_x\si_n^y(\si_{n+1}^x+\si_{n-1}^x)
                      +J_y\si_n^x(\si_{n+1}^y+\si_{n-1}^y),\\
                       & &(n=2,\cdots,N-1) \nonumber
\ea
\ba
\frac{d}{dt} \si_{1}^x &=& -J_y\si_1^z\si_2^y+J_z\si_1^y\si_2^z
                       +{\rm  sn} \eta(-2A_{-}\si_1^y+i(B_{-}-C_{-})\si_1^z),\\
\frac{d}{dt} \si_{1}^y &=& -J_z\si_1^x\si_2^z+J_x\si_1^z\si_2^x
                       +{\rm  sn} \eta(2A_{-}\si_1^x-(B_{-}+C_{-})\si_1^z),\\
\frac{d}{dt} \si_{1}^z &=& -J_x\si_1^y\si_2^x+J_y\si_1^x\si_2^y
                       -2i{\rm  sn} \eta(B_{-}\si_1^{+}-C_{-}\si_1^{-}),
\ea
\ba
\frac{d}{dt} \si_{N}^x &=& -J_y\si_{N-1}^y\si_N^z+J_z\si_{N-1}^z\si_N^y
                       +{\rm  sn} \eta(-2A_{+}\si_N^y+i(B_{+}-C_{+})\si_N^z),\\
\frac{d}{dt} \si_{N}^y &=& -J_z\si_{N-1}^z\si_N^x+J_x\si_{N-1}^x\si_N^z
                       +{\rm  sn} \eta(2A_{+}\si_N^x-(B_{+}+C_{+})\si_N^z),\\
\frac{d}{dt} \si_{N}^z &=& -J_x\si_{N-1}^x\si_N^y+J_y\si_{N-1}^y\si_N^x
                       -2i{\rm  sn} \eta(B_{+}\si_N^{+}-C_{+}\si_N^{-}).
\ea
In order to rewrite equations above in the Lax form, we consider the following 
operator version of an auxiliary linear problem,
\ba
\Psi_{n+1} &=& L_n(u)\Psi_n , \;\;\;(n=1,2,\cdots,N) \\
\frac{d}{dt}\Psi_n &=& M_n(u)\Psi_n , \;\;\;(n=2,\cdots,N) \\
\frac{d}{dt}\Psi_{N+1} &=& Q_N(u)\Psi_{N+1}, \\
\frac{d}{dt}\Psi_1 &=& Q_1(u)\Psi_1, 
\ea
where $u$ is the spectral parameter which does not depend on the time $t$,
$L_n(u), M_n(u), Q_1(u)$ and $Q_N(u)$ are called the Lax pair. $Q_1(u)$ and
$Q_N(u)$ are responsible for the boundary conditions. The consistency conditions for 
the equations (12)-(15) are  the following Lax equations:
\ba
\frac{d}{dt}L_n(u) &=& M_{n+1}(u)L_n(u)-L_n(u)M_n(u), \;\;\; (n=2,\cdots,N-1)\\ 
\frac{d}{dt}L_N(u) &=& Q_N(u)L_N(u)-L_N(u)M_N(u),\\
\frac{d}{dt}L_1(u) &=& M_2(u)L_1(u)-L_n(u)Q_1(u).
\ea
From the eqs. (3)-(5) and (16), we see that they are the same as those in the case of periodic 
boundary condition. For the $XYZ$ model with periodic boundary condition, the 
Lax pair $L_n, M_n$  has been given by Sogo and Wadati\cite{sogo}. So, for the 
open $XYZ$ model, we only have to construct operators $Q_1(u)$ and $Q_N(u)$. 
If we write $Q_1(u)$ and $Q_N(u)$ in the following forms:
\be
Q_1(u)=(a_{ij}^{(-)})_{2\times 2},\;\;\; Q_N(u)=(a_{ij}^{(+)})_{2\times 2}
\ee
and
\be
a_{ij}^{(-)} = \sum_{l=0}^{3}a_{ij}^{(-)l}\si_1^l , \;\;\;\;
a_{ij}^{(+)} = \sum_{l=0}^{3}a_{ij}^{(+)l}\si_N^l ,
\ee
where $\si^0=1$, $\si^1, \si^2, \si^3=\si^x, \si^y, \si^z$ are the Pauli matrices, 
then after a lengthy calculations, from eqs. (17) and (18), we obtain the expressions
for $a_{ij}^{(\pm)l} (i,j=1,2, l=0,1,2,3)$  as follows,
\ba
a_{11}^{(\pm)0} &=& \frac{4i{\rm  sn}  \eta A_{\pm}}{D} a(u)b(u);\\
a_{11}^{(\pm)1} &=& \frac{i{\rm  sn}  \eta }{D}\{ \pm(B_{\pm}-C_{\pm})c(u)(-2e(u)+f(u)+g(u)) \nonumber \\
                & &  -(B_{\pm}+C_{\pm})(b^2(u)+c^2(u)-f^2(u))\}; \\  
a_{11}^{(\pm)2} &=& \pm\frac{{\rm  sn}  \eta }{D}\{ \pm(B_{\pm}-C_{\pm})(a^2(u)+c^2(u)-e^2(u))\nonumber \\
                & &  +(B_{\pm}+C_{\pm})c(u)(f(u)-g(u))\}; \\
a_{11}^{(\pm)3} &=& G_3-\frac{i{\rm  sn} \eta A_{\pm}}{D}d(u);\\
a_{12}^{(\pm)0} &=& \frac{2i{\rm  sn} \eta }{D} c(u)\{(B_{\pm}-C_{\pm})a(u)+(B_{\pm}+C_{\pm})b(u)\};\\
a_{12}^{(\pm)1} &=& G_1\pm\frac{2i{\rm  sn} \eta A_{\pm}}{D} b(u)(-2e(u)+f(u)+g(u));\\
a_{12}^{(\pm)2} &=&-iG_2\pm \frac{2{\rm  sn} \eta A_{\pm}}{D} a(u)(-f(u)+g(u));\\
a_{12}^{(\pm)3} &=&\mp \frac{i{\rm  sn} \eta }{D}(f(u)+g(u))\{(B_{\pm}-C_{\pm})a(u)
                    +(B_{\pm}+C_{\pm})b(u)\};\\
a_{21}^{(\pm)0} &=&\frac{2i {\rm  sn} \eta }{D}c(u)\{-(B_{\pm}-C_{\pm})a(u)
                    +(B_{\pm}+C_{\pm})b(u)\};\\
a_{21}^{(\pm)1} &=&G_1\mp\frac{2i {\rm  sn} \eta A_{\pm} }{D}b(u)(-2e(u)+f(u)+g(u));\\
a_{21}^{(\pm)2} &=&i G_2\pm\frac{2 {\rm  sn} \eta A_{\pm} }{D}a(u)(-f(u)+g(u));\\
a_{21}^{(\pm)3} &=&\pm\frac{i {\rm  sn} \eta }{D}(f(u)+g(u))\{-(B_{\pm}-C_{\pm})a(u)
                    +(B_{\pm}+C_{\pm})b(u)\};  \\
a_{22}^{(\pm)0} &=&-a_{11}^{(\pm)0} ;\\
a_{22}^{(\pm)1} &=& \frac{i {\rm  sn} \eta }{D}\{ \pm(B_{\pm}-C_{\pm})c(u)(2e(u)-f(u)-g(u)) \nonumber \\
                & &  -(B_{\pm}+C_{\pm})(b^2(u)+c^2(u)-f^2(u))\}; \\
a_{22}^{(\pm)2} &=&\frac{{\rm  sn} \eta }{D}\{ (B_{\pm}-C_{\pm})(a^2(u)+c^2(u)-e^2(u))\nonumber \\
                & &  \mp(B_{\pm}+C_{\pm})c(u)(f(u)-g(u))\}; \\
a_{22}^{(\pm)3} &=&-G_3-\frac{i{\rm  sn} \eta A_{\pm}}{D}d(u)
\ea
where
\ba
a(u) &=& w_2 w_3+w_1w_4,\;\;\; b(u)=w_1 w_3+w_2w_4,\nonumber \\
c(u) &=& w_1 w_2+w_3w_4, \nonumber \\
d(u) &=&(w_2-w_1)^2\{(w_3-w_4)^2-(w_2+w_1)^2\}+ \nonumber\\
     & &   +(w_2+w_1)^2\{(w_3+w_4)^2-(w_2-w_1)^2\},\\
e(u) &=& w_1^2-w_3^2,\;\;\; f(u)=w_2^2-w_3^2,\;\;\;g(u)=w_1^2-w_4^2,\nonumber\\
w_1(u)+w_2(u)&=&{\rm  sn} \eta,\;\;\;w_1(u)-w_2(u)=k {\rm  sn} \eta {\rm  sn} u {\rm  sn}(u+ \eta),\nonumber\\
w_3(u)+w_4&=&{\rm  sn}(u+ \eta),\;\;\;w_4(u)-w_3={\rm  sn} u,\nonumber
\ea
for the explicit expressions of $D, G_1,G_2,G_3$ see ref. \cite{sogo}. In the 
trigonometric limit $k\rightarrow 0$, where ${\rm  sn} u\rightarrow \sin u$ and taking
$J_x=J_y=1, J_z=\cos \eta, B_{\pm}=C_{\pm}=0$, we recover the result in the case 
of open $XXZ$ model given in the ref. \cite{zhou}(up to a replacement of $\eta$ by $2\eta$).\\

The eqs.(19)-(36) combined with $L_n, M_n$ given in ref. \cite {sogo} are the Lax pair for the
open $XYZ$ spin-chain. Now we make two remarks. For systems with periodic boundary conditions, 
Korepin et al. \cite{kor} proved that the Lax equation can follow from the Yang-Baxter equation. 
In the case of open boundary conditions, the integrability is  guaranteed by both the 
Yang-Baxter equation and  the reflection equations. Then we should discuss the relations
between the Lax representation and the reflection equations . Furthermore, if we construct 
conserved quantities from the transfer matrix, then there are some constraints for $Q_1(u),
 Q_N(u)$ and $K^{\pm}$-matrices. Whether the constraints hold for $Q_1(u), Q_N(u)$ given in 
 this letter and $K^{\pm}$-matrices in ref. \cite{ina} should be considered. We expect to
 discuss these problems elsewhere.\\
 
 We would like to thank Prof. Ke Wu and Prof. Shi-kun Wang for useful discussions.

\end{document}